\documentclass[lettersize,journal]{IEEEtran}
\usepackage{amsmath,amsfonts}
\usepackage{amsfonts,amsthm,bm}
\usepackage{algorithmic}
\usepackage{xcolor}
\usepackage{algorithm}
\usepackage{array}
\usepackage[caption=false,font=normalsize,labelfont=sf,textfont=sf]{subfig}
\usepackage{textcomp}
\usepackage{stfloats}
\usepackage{url}
\usepackage{verbatim}
\usepackage{graphicx}
\usepackage{tabularx}
\usepackage{booktabs}       
\usepackage{cite}
\usepackage{comment}
\usepackage[capitalise]{cleveref}
\usepackage{csquotes}
\usepackage{pgfplots}
\usepackage{siunitx}
\DeclareSIUnit\MW{\mega\watt}
\usetikzlibrary{quotes,angles, shapes.symbols}
\usetikzlibrary{calc}

\usepackage[acronym,toc,nonumberlist]{glossaries}
\newacronym{IML}{IML}{Interpretable Machine Learning}
\newtheorem{theorem}{Theorem}

\newacronym{ML}{ML}{Machine Learning}
\newacronym{SHAP}{SHAP}{SHapley Additive exPlanations}
\newacronym{RES}{RES}{Renewable Energy Sources}
\newacronym{PFI}{PFI}{Permutation Feature Importance}
\newacronym{DCPF}{DCPF}{DC Power Flow}
\newacronym{PTDF}{PTDF}{Power Transfer Distribution Factors}
\newacronym{XGBoost}{XGBoost}{eXtreme Gradient Boosting}

\begin{document}

\title{Interpretable Machine Learning for Power Systems: Establishing Confidence in SHapley Additive exPlanations}

\author{Robert I. Hamilton,~{\textit{Member, IEEE}}, Jochen Stiasny, Tabia Ahmad,~{\textit{Member, IEEE}}, Samuel Chevalier,~{\textit{Member, IEEE}}, Rahul Nellikkath, Ilgiz Murzakhanov,~{\textit{Student Member, IEEE}}, Spyros Chatzivasileiadis,~{\textit{Senior Member, IEEE}} and Panagiotis N. Papadopoulos,~{\textit{Member, IEEE}}
\thanks{R. I. Hamilton, T. Ahmad and P. N. Papadopoulos are with the Department of Electronic and Electrical Engineering at the University of Strathclyde, Glasgow, Scotland and supported by the UKRI Future Leaders Fellowship MR/S034420/1 (e-mail: robert.hamilton@strath.ac.uk). For the purpose of open access, the authors have applied for a Creative Commons Attribution (CC BY) license to any Author Accepted Manuscript version arising from this submission.

Jochen Stiasny, Rahul Nellikkath, Ilgiz Murzakhanov and Spyros Chatzivasileiadis are supported by the ID-EDGe project, funded by Innovation Fund Denmark, Grant Agreement No. 8127-00017B, and by the ERC Starting Grant VeriPhIED, Grant Agreement No. 949899. Samuel Chevalier is supported by the HORIZON-MSCA-2021 Postdoctoral Fellowship Program, Project \#101066991 – TRUST-ML.}

}




\maketitle

\begin{abstract}
Interpretable Machine Learning (IML) is expected to remove significant barriers for the application of Machine Learning (ML) algorithms in power systems. This letter first seeks to showcase the benefits of SHapley Additive exPlanations (SHAP) for understanding the outcomes of ML models, which are increasingly being used. Second, we seek to demonstrate that SHAP explanations are able to capture the underlying physics of the power system. To do so, we demonstrate that the Power Transfer Distribution Factors (PTDF)\textemdash a physics-based linear sensitivity index\textemdash can be derived from the SHAP values. To do so, we take the derivatives of SHAP values from a ML model trained to learn line flows from generator power injections, using a simple DC power flow case in the 9-bus 3-generator test network. In demonstrating that SHAP values can be related back to the physics that underpin the power system, we build confidence in the explanations SHAP can offer.


\end{abstract}



\section{Introduction}
\IEEEPARstart{P}{ower} systems are extremely complex high dimensional systems, the complexity of which is only set to increase with the connection of renewable energy sources and inclusion of other energy vectors such as heating and transportation. Understanding complex power system phenomena in this context is crucial for ensuring continued reliable power supply. In recent years, many \gls{ML} applications to predict the behaviour of various aspects of power systems have been developed\textemdash some of which are summarised in \cite{ozcanli2020deep}. While these black-box \gls{ML} algorithms have shown good accuracy and computational savings, their applicability to mission-critical infrastructure such as power systems is limited due to absence of trustworthy explanations. 


The premise of \emph{interpretable} \gls{ML}\textemdash an emerging area of research\textemdash is to provide detailed explanations of \gls{ML} model predictions in order to enhance confidence in the model predictions. Among the post-hoc \gls{ML} model interpretability methods, which \cite{molnar2020interpretable} presents concisely, \gls{SHAP} has gained some initial traction in power systems, as \cite{Machlev:2022} reviews. The focus of the limited number of applications in literature, however, usually centre around applying \gls{SHAP}, rather than analysing the method itself.

\gls{SHAP} \cite{lundberg2017unified} is model agnostic (i.e., can be applied to different types of \gls{ML} models) and is of the class of additive feature attribution methods; meaning that it attributes an effect of a feature $x_i$ on the prediction of a model $f(\bm{x})$. 


Such methods construct a simple additive \textit{explanation model}, $g$\textemdash which is a linear function of binary variables\textemdash to represent the complex \textit{original model}, $f$. In the \gls{SHAP} framework, the explanation model is expressed as a \enquote{conditional expectation function of the original model} \cite{lundberg2017unified}. Simplified inputs $x'$ are used to map to the original input through mapping function $h_x$, where $x=h_{x} (x')$; ensuring $g(z')\approx g(h_x(z'))$, whenever $z'\approx x'$ and where $z'\in \{0,1\}^M$ and $M$ is the number of simplified input features. Thus an effect $\phi_i$ (where $\phi_i \in \mathbb{R}$) is attributed to each feature, the sum of which approximates $f(\bm{x})$ as per (\ref{eq: additive}). 
\begin{align}
    g(z') = \phi_0 + \textstyle\sum_{i=1}^{M} {\phi_i z'_i} \label{eq: additive}
\end{align}
This quantification is based on Shapley values \cite{shapley1953stochastic}, which stems from game theory and describes the average marginal contribution of a player to all coalitions in which the player contributes.
The \gls{SHAP} framework often incorporates approximate \gls{SHAP} values for computational efficiency \cite{lundberg2017unified}.  \gls{SHAP} offers local explanations for a single point, which can be repeated for multiple points to achieve global interpretations. For more details, the interested reader can refer to \cite{lundberg2020local}.



This letter initially showcases \gls{SHAP} interpretations as a tool for understanding power system \gls{ML} models using a simple case study (Section II), and then establishes the capability of \gls{SHAP} explanations to capture underlying physics of the power system\textemdash thus enhancing confidence in \gls{SHAP} as an interpretability method (Section III). We achieve this by deriving \gls{PTDF} (i.e., the sensitivity of line flows to power injections) from \gls{SHAP} values. 
In doing so, we seek to build confidence in \gls{SHAP} as a tool for interpreting \gls{ML} models. 

\section{Using SHAP Values to Interpret ML Models for Line Flow Prediction: 9-Bus Case Study}
\label{results}

\begin{figure}[t]
	\centering
	\includegraphics[scale=0.35]{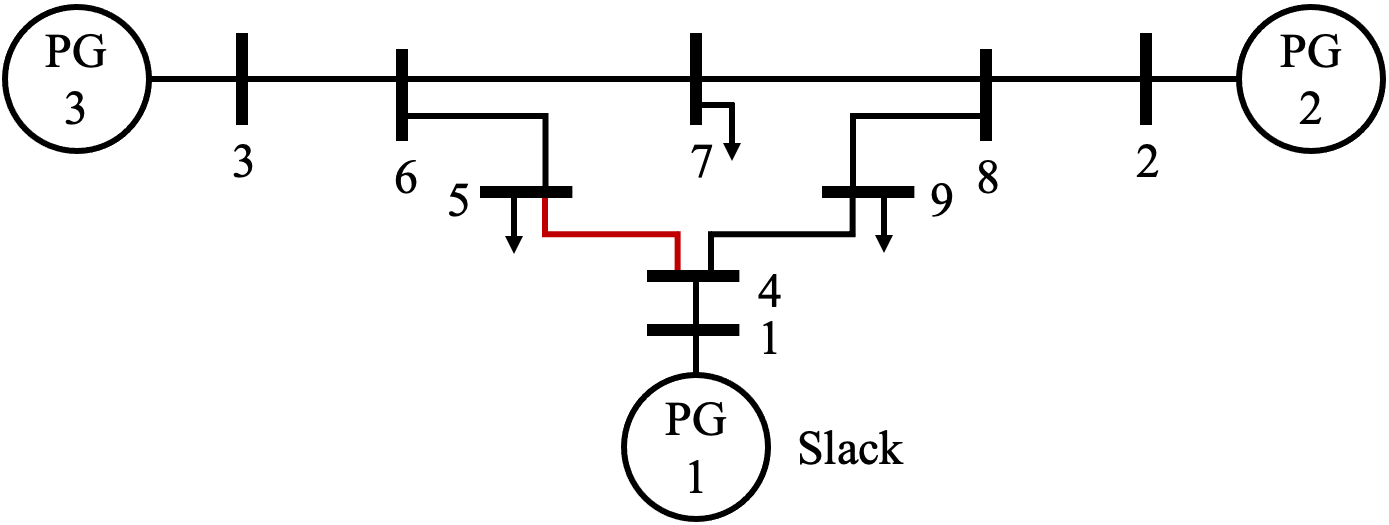}
	\caption[9-bus 3-generator test network with line 4-5 highlighted for analysis.]{9-bus 3-generator test network with line 4-5 highlighted for analysis.}
        \label{fig:3-generator_network}
	\end{figure}

Individual \gls{XGBoost} \cite{chen2016xgboost} regression models $f_{line,i-j}(x)$ are trained to predict the active power flow ($F_{line,i-j}$) on each line of the the 9-bus 3-generator test network (\cref{fig:3-generator_network}). This is achieved using a set of input features $x$, $x \in \bm{P}$ for a simple DC power flow case, with a 75-25\% train-test split. The parameters of the test network are taken from Matpower \cite{zimmerman1997matpower}. For the trained models, \gls{SHAP} values are calculated (using the framework set out in \cite{lundberg2017unified}, implemented in Python).


\subsection{Database Creation}
To generate the database upon which the \gls{ML} models are trained, samples of $PG2$ and $PG3$ are drawn independently from a uniform distribution in the range [0, 500] \si{MW} for a total of 1001 generation scenarios. For the sake of this case study, but without loss of generality, demand is assumed constant at 315 \si{MW}. $PG1$ acts as the slack bus. Following the execution of a DC power flow, the active power flow on each line is recorded ($F_{line,i-j}$) in the database. Since $PG1$ is the slack generator (and therefore not an independent feature), it is excluded from the database. The final database contains $PG2$ and $PG3$ as training features and $F_{line,i-j}$ as targets for all 1001 operational scenarios. Database available in \cite{Hamilton_GitHub}.








\subsection{SHAP Insights: Analysis of $f_{line,4-5}(x)$}
To gain insights into each \gls{ML} model, the Tree \gls{SHAP} algorithm  (a polynomial time algorithm that leverages the tree structure \cite{lundberg2020local}) is applied to each model. Analysis of a single local explanation and a global model interpretation of $f_{line,4-5}(x)$ is presented below. Similar analysis can be extracted for the remaining \gls{ML} models $f_{line,i-j}(x)$.
\subsubsection{Local Explanation}
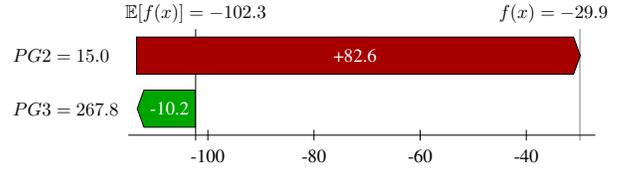
\begin{figure}[ht]
		\centering
		\resizebox{\linewidth}{!}{%
		\def\SHAPzero{-10.23cm}
\def\SHAPone{1.02cm}
\def\SHAPtwo{8.36cm}
\def\functionValue{-2.99cm}
\def\generatorValue{-13.8cm}
\def\xticksHeight{0.1cm}
\def\arrowHeight{0.7cm}

\begin{tikzpicture}[]%
    \node (origin) [] at (\SHAPzero, 0) {};
  \draw (\SHAPzero,-0.1cm) -- (\SHAPzero,2cm) node[yshift=0.3cm] {$\mathbb{E}[f(x)] = -102.3$};
    \draw[-, gray] (\functionValue,-0.1cm) -- (\functionValue,2cm) node [yshift=0.3cm, xshift=-0.5cm, black] {$f(x) =-29.9$};
    \draw (-4cm,\xticksHeight) -- (-4cm,-\xticksHeight) node[yshift=-0.3cm] {-40};
    \draw (-6cm,\xticksHeight) -- (-6cm,-\xticksHeight) node[yshift=-0.3cm] {-60};
    \draw (-8cm,\xticksHeight) -- (-8cm,-\xticksHeight) node[yshift=-0.3cm] {-80};
    \draw (-10cm,\xticksHeight) -- (-10cm,-\xticksHeight) node[yshift=-0.3cm] {-100};
  \draw[-] (-11.5cm,0) -- (-2.7cm,0) node [] {};
  \node[signal, draw, fill=green!65!black, signal to=west, anchor=east, minimum width=\SHAPone, signal pointer angle=140, minimum height=\arrowHeight, text=white] at ($(origin) + (0,0.5)$) {-10.2};
  \node[signal, draw, fill=red!65!black, signal to=east, minimum height=\arrowHeight, minimum width=\SHAPtwo, anchor=west, signal pointer angle=140, text=white] at ($(origin) + (-\SHAPone,1.5) + (-0.1, 0.0)$) {+82.6};
  \node[right] at (\generatorValue, 0.5) {$PG3=267.8$};
  \node[right] at (\generatorValue, 1.5) {$PG2=15.0$};
\end{tikzpicture}
		}%
		\caption[]{Local SHAP explanation for $F_{line,4-5}$ showing how SHAP values of feature $\phi_i$ impact the move from model expectation $\mathbb{E}[f(x)]$ to final prediction $f(x)$ for a single operational scenario.}
        \label{fig:line45_local_summary}
	\end{figure}

Local explanations provide the contributions of features (i.e. \gls{SHAP} values, $\phi_i$) in shifting the model prediction from the model expectation ($\mathbb{E}[f(x)]$)\textemdash that would be predicted if we had no feature information\textemdash to the final prediction ($f(x)$) for a single operational scenario. Therefore the \gls{SHAP} values ($\phi_i$) are given in the units of the variable being predicted. In this study, features are the $PG2$ and $PG3$ injections. 
An example of a local \gls{SHAP} explanation for the model $f_{line,4-5}(x)$ is given in Fig. \ref{fig:line45_local_summary}. For this particular operational scenario, the baseline model expectation ($\mathbb{E}[f(x)]$) for $F_{line,4-5}$ is \SI{-102.3}{\MW}. $PG3$ setpoint is \SI{267.8}{\MW}, which decreases the prediction by \SI{-10.2}{\MW} ($\phi_{PG3}$)) from $\mathbb{E}[f(x)]$ to \SI{-112.5}{\MW}. The setpoint for $PG2$ is \SI{15.0}{\MW}, which increases the model prediction by \SI{82.6}{\MW} ($\phi_{PG2}$). This results in the final prediction of $f(x)= \SI{-29.9}{\MW}$. This is consistent with (\ref{eq: additive}).
Since $\lvert\phi_{PG2}\rvert > \lvert\phi_{PG3}\rvert$, $PG2$ has a greater effect in this scenario and thus is placed higher on the y-axis.

\subsubsection{Global Interpretation}
Global interpretations comprise of the local explanations for the entire training database, making them consistent. Analysis of \gls{SHAP} values in the global frame assists in understanding the global model structure. 

\begin{figure}[ht]
		\centering
		\includegraphics[scale=0.5]{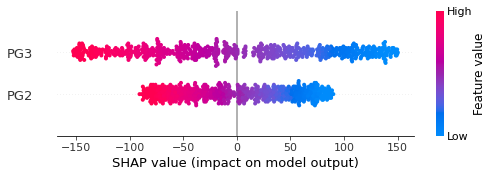}
		\caption[]{Global SHAP interpretation for $F_{line,4-5}$. Each point for each feature is an operational scenario in the training dataset.}
        \label{fig:line45_global_summary}
	\end{figure}

The global \gls{SHAP} plot (Fig. \ref{fig:line45_global_summary}) gives the \gls{SHAP} value for each feature (x-axis) with respect to the feature value (color-axis). Features are ordered on the y-axis based on importance (defined here as the mean of all \gls{SHAP} values for all operational scenarios). In this case, $PG3$ is found to have a higher importance than $PG2$ (note: this is the inverse to the local explanation given in Fig. \ref{fig:line45_local_summary} for that particular case, indicating how local sensitivity vs. global importance need not necessarily be the same). It can be seen that as $PG3$ increases, the \gls{SHAP} values ($\phi_{PG3}$) decrease from positive to negative. This means that as $PG3$ increases the impact it has on $F_{line,4-5}$ goes from increasing the baseline prediction, to decreasing it. This is consistent with the theory of power flow when observing Fig. \ref{fig:3-generator_network}\textemdash where one would expect a higher $PG3$ setpoint to decrease $F_{line,4-5}$ in this simple DC power flow example. A similar trend can be seen for the \gls{SHAP} values for $PG2$ ($\phi_{PG2}$)\textemdash although the impact is smaller than $PG3$, indicated by a smaller spread of \gls{SHAP} values on the x-axis. The feature value is given on the color axis and is normalised based on the min/maximum feature value for trend identification, however the actual feature value (in \si{MW}) can also be extracted. This showcases how \gls{SHAP} values can give some interesting insights about how power system features, here the power injections, affect our outputs of interest, here the line flows. This can be especially useful in more complex cases, where analytical relationships are difficult to extract.
\section{Derivation of PTDF from SHAP Values}
\label{SHAP_PTDF_Section}
To show the link between \gls{SHAP} values and the \gls{PTDF}, we consider $f_{line,i-j}(x)$ to be a linear statistical model $f(\bm{x}) = \bm{w}^\top \bm{x} + b$ whose features ($\bm{x} \in {\bm P}$) are assumed to be independent. Here the features are the power injections $PG2$ and $PG3$.

\begin{theorem}\label{eq: Th1}
The derivatives of the \gls{SHAP} values $\phi_i(f,{\bm x})$, $i\ne 0$, associated with $f(\bm{x})$ yield exactly the \gls{PTDF} of this network.
\begin{proof}
Using Corollary 1 from \cite{lundberg2017unified}, the \gls{SHAP} values $\phi_i(f,{\bm x})$ associated with $f(\bm{x})$ are given by
\begin{align}
    \phi_0 (f, {\bm x}) &= b\\
    \phi_i(f, {\bm x}) &= \bm{w}_i (\bm{x}_i - \mathbb{E}[{\bm{X}_i}]), \; i\ne 0 \label{eq: phi_i}
\end{align}
where $\bm{X}_i$ is the training data associated with the $i^{\rm th}$ feature. Perturbing the $i^{\rm th}$ feature (i.e., continuous regressor) yields
\begin{align}\label{eq: derivative}
\frac{\partial\phi_{i}}{\partial\bm{x}_{i}}=\bm{w}_{i}.
\end{align}
Since $\bm{w}_{i}$ relates the sensitivity of line flow to the $i^{\rm th}$ injection, the \gls{SHAP} derivative is equivalent to a \gls{PTDF}. 
\end{proof}
\end{theorem}
For a definition and analytical derivation of the \gls{PTDF}, the interested reader can refer to \cite{DBLP:journals/corr/abs-1811-00943}. Strictly speaking, the result of \cref{eq: Th1} is only valid when the underlying statistical model is linear (or affine). However, \gls{ML} practitioners often use models which have the capacity to be aggressively \textit{nonlinear}. \gls{SHAP} can be applied in either case and 
as the trained models effectively still behave like \textit{linear} models, 
(\ref{eq: phi_i})-(\ref{eq: derivative}) will remain approximately valid. To show this experimentally, we note that the sum across all \gls{SHAP} values should yield a model with linear sensitivity to power injections, i.e., an approximate \gls{PTDF} vector $\hat{\bm{D}}$, and a constant offset term $\epsilon$:
\begin{align}
\sum_{i}\phi_{i}(f,x)=\bm{w}^{\top}(\bm{x}-\mathbb{E}[\bm{X}])+b\approx\hat{\bm{D}}^{\top}{\bm P}+\epsilon.
\end{align}
By collecting a library of \gls{SHAP} values $\Phi$ associated with a library of sampled injection values $\bf P$, a regression procedure can yield the \gls{PTDF} approximation $\hat{{\bm D}}$:
\begin{align}\label{eq: reg}
\left[\begin{array}{c}
\hat{{\bm D}}\\
\epsilon
\end{array}\right]={\bf P}{}^{+}\Phi,
\end{align}
where $(\cdot)^{+}$ denotes Moore–Penrose pseudoinversion. 
Table~\ref{tab:SHAP-PTDF} presents the analytical \gls{PTDF} and the \gls{SHAP}-based \gls{PTDF} $\hat{\bm{D}}$ for the relevant buses for the case study in Section II. These experimental results support \cref{eq: Th1}, showing that the derivatives of the \gls{SHAP} values are equivalent to the \gls{PTDF} for this network.



\begin{table}[ht]
  \caption{9-Bus 3-Generator Test Network PTDF \& SHAP.}
  \label{tab:SHAP-PTDF}
  \centering
  \begin{tabular}{ c|cc|cc }
    \toprule
    &  \multicolumn{2}{c|}{True physical PTDF, $\bm{D}$} & \multicolumn{2}{c}{SHAP-based PTDF, $\hat{\bm{D}}$}\\
    \midrule
     Line  & Bus 2 & Bus 3  & Bus 2 & Bus 3\\
    \midrule
 Line 1-4  & -1.0000 & -1.0000  & -0.9999 & -0.9999 \\ 
 Line 4-5  & -0.3613 & -0.6152  & -0.3613 & -0.6151 \\
 Line 5-6  & -0.3613 & -0.6152  & -0.3613 & -0.6151 \\ 
 Line 3-6  & 0 & 1.0000  & 0.0000 & 0.9999 \\ 
 Line 6-7  & -0.3613 & 0.3848  & -0.3613 & 0.3848 \\ 
 Line 7-8  & -0.3613 & 0.3848  & -0.3613 & 0.3848 \\ 
 Line 8-2  & -1.0000 & 0  & -1.0000 & 0.0000 \\ 
 Line 8-9  & 0.6387 & 0.3848  & 0.6386 & 0.3848 \\ 
 Line 9-4  & 0.6387 & 0.3848  & 0.6386 & 0.3848 \\
    \bottomrule
  \end{tabular}
\end{table}


\section{Potential Applications of SHAP Framework}
The main motivations for using \gls{ML} are its capacity to provide explicit mappings of complex functions and accelerate computationally heavy tasks. One could use such a \gls{ML} model directly, however they are often black-box and therefore difficult to interpret. To build confidence in the \gls{ML} model and foster widespread use in the power system domain, understanding how the model makes predictions and identify potential relationships with power system physics is essential. Model interpretability techniques can be used to achieve this to either bolster existing knowledge, or infer new information about the power system. Ideally this should be standardised across the industry.

We believe that \gls{SHAP} has the potential to be such a technique. This arises from the ability to look at local explanations, e.g., for a specific operating point, as well as global explanations, i.e., ranking feature importance for a model. Based on these explanations, we could (i) compare the performance of different \gls{ML} models beyond pure accuracy as well as investigate their implicit biases, (ii) search for simpler operational rules derived from a \gls{ML} model, (iii) improve the dataset generation by focusing on contradicting explanations. 

For these processes to be effective, it is important to understand the characteristics of the explanation method, i.e., \gls{SHAP}, itself. In this context, the ability to link \gls{SHAP} and its explanations to physics (e.g., \gls{PTDF}) is a desirable property. For a more complex case than the presented one, such links could reveal interesting insights and also help to gain confidence in both \gls{ML} models and indeed \gls{SHAP} as an interpretability method for power systems.   

\section{Conclusion}
Interpretable Machine Learning (IML) in power systems will become necessary to understand increasingly complex Machine Learning (ML) models used in academia and industry. For wide-spread adoption, confidence must be built in the interpretation method. Physical equivalence is one such way of developing confidence, which we have shown in this letter for SHapley Additive exPlanations (SHAP) and Power Transfer Distribution Factors (PTDF) in a simple DC power flow case. Extending from this linear case to more complex nonlinear problems will likely be a fruitful avenue of future research. 

\bibliographystyle{IEEEtran}
\bibliography{references}

\begin{thebibliography}{10}
\providecommand{\url}[1]{#1}
\csname url@samestyle\endcsname
\providecommand{\newblock}{\relax}
\providecommand{\bibinfo}[2]{#2}
\providecommand{\BIBentrySTDinterwordspacing}{\spaceskip=0pt\relax}
\providecommand{\BIBentryALTinterwordstretchfactor}{4}
\providecommand{\BIBentryALTinterwordspacing}{\spaceskip=\fontdimen2\font plus
\BIBentryALTinterwordstretchfactor\fontdimen3\font minus
  \fontdimen4\font\relax}
\providecommand{\BIBforeignlanguage}[2]{{%
\expandafter\ifx\csname l@#1\endcsname\relax
\typeout{** WARNING: IEEEtran.bst: No hyphenation pattern has been}%
\typeout{** loaded for the language `#1'. Using the pattern for}%
\typeout{** the default language instead.}%
\else
\language=\csname l@#1\endcsname
\fi
#2}}
\providecommand{\BIBdecl}{\relax}
\BIBdecl

\bibitem{ozcanli2020deep}
A.~K. Ozcanli, F.~Yaprakdal, and M.~Baysal, ``Deep learning methods and
  applications for electrical power systems: A comprehensive review,''
  \emph{International Journal of Energy Research}, vol.~44, no.~9, pp.
  7136--7157, 2020.

\bibitem{molnar2020interpretable}
C.~Molnar, \emph{Interpretable machine learning}.\hskip 1em plus 0.5em minus
  0.4em\relax Lulu. com, 2020.

\bibitem{Machlev:2022}
R.~Machlev, L.~Heistrene, M.~Perl, K.~Levy, J.~Belikov, S.~Mannor, and
  Y.~Levron, ``Explainable artificial intelligence (xai) techniques for energy
  and power systems: Review, challenges and opportunities,'' \emph{Energy and
  AI}, vol.~9, p. 100169, 2022.

\bibitem{lundberg2017unified}
S.~M. Lundberg and S.-I. Lee, ``A unified approach to interpreting model
  predictions,'' \emph{Advances in neural information processing systems},
  vol.~30, 2017.

\bibitem{shapley1953stochastic}
L.~S. Shapley, ``Stochastic games,'' \emph{Proceedings of the national academy
  of sciences}, vol.~39, no.~10, pp. 1095--1100, 1953.

\bibitem{lundberg2020local}
S.~M. Lundberg, G.~Erion, H.~Chen, A.~DeGrave, J.~M. Prutkin, B.~Nair, R.~Katz,
  J.~Himmelfarb, N.~Bansal, and S.-I. Lee, ``From local explanations to global
  understanding with explainable ai for trees,'' \emph{Nature machine
  intelligence}, vol.~2, no.~1, pp. 56--67, 2020.

\bibitem{chen2016xgboost}
T.~Chen and C.~Guestrin, ``Xgboost: a scalable tree boosting system acm sigkdd
  international conference on knowledge discovery and data mining,''
  \emph{ACM}, pp. 785--794, 2016.

\bibitem{zimmerman1997matpower}
R.~D. Zimmerman, C.~E. Murillo-S{\'a}nchez, and D.~Gan, ``Matpower,''
  \emph{PSERC.[Online]. Software Available at: http://www. pserc. cornell.
  edu/matpower}, 1997.

\bibitem{Hamilton_GitHub}
\BIBentryALTinterwordspacing
R.~Hamilton, ``{SHAP Database Repository},'' 2022. [Online]. Available:
  \url{https://github.com/RobertIHamilton/SHAP_database}
\BIBentrySTDinterwordspacing

\bibitem{DBLP:journals/corr/abs-1811-00943}
\BIBentryALTinterwordspacing
S.~Chatzivasileiadis, ``Lecture notes on optimal power flow {(OPF)},''
  \emph{CoRR}, vol. abs/1811.00943, 2018. [Online]. Available:
  \url{http://arxiv.org/abs/1811.00943}
\BIBentrySTDinterwordspacing

\end{thebibliography}

\newpage
\vfill
\end{document}